\documentclass{pasa}%

\usepackage{graphicx}
\usepackage{subcaption}
\usepackage{amsmath}

\title[CRRLs in Orion Molecular Cloud]{Low-Frequency Carbon Recombination Lines in the Orion Molecular Cloud Complex}

\author[Tremblay et al.]{Chenoa D. Tremblay$^{1}$, Christopher H. Jordan$^{1,2}$, Maria Cunningham$^{3}$, Paul A. Jones$^{3}$ and Natasha~Hurley-Walker$^{1}$
\affil{$^1$International Centre for Radio Astronomy Research, Curtin University, 1 Turner Ave Bentley, WA 6155}%
\affil{$^2$ARC Centre of Excellence for All-sky Astrophysics (CAASTRO)}
\affil{$^3$School of Physics, University of New South Wales, Sydney, NSW 2052, Australia}
}%

\jid{PASA}
\doi{10.1017/pas.\the\year.xxx}
\jyear{\the\year}

\usepackage{aas_macros}
\usepackage{hyperref} 
\hypersetup{colorlinks,citecolor=blue,linkcolor=blue,urlcolor=blue}

\hypersetup{draft}

\begin{document}

\begin{frontmatter}
\maketitle

\begin{abstract}
We detail tentative detections of low-frequency carbon radio recombination lines from within the Orion molecular cloud complex observed at 99--129\,MHz. These tentative detections include one alpha transition and one beta transition over three locations and are located within the diffuse regions of dust observed in the infrared at 100\,$\mu$m, the H$\alpha$ emission detected in the optical, and the synchrotron radiation observed in the radio. With these observations, we are able to study the radiation mechanism transition from collisionally pumped to radiatively pumped within the H{\sc ii} regions within the Orion molecular cloud complex.   
\end{abstract}

\end{frontmatter}

\section{INTRODUCTION }
\label{sec:intro}
Radio recombination lines (RRLs) are produced in regions of diffuse partially ionised gas, typically associated with interstellar radiation fields from stars \citep{Sorochenko-2010}.  Atoms in high quantum states are important tracers of ionised gas within photon-dominated regions (PDRs), where the far-ultraviolet (FUV) radiation from stars that are embedded within the molecular clouds interact with the neutral hydrogen gas surrounding the H{\sc ii} regions \citep{Andree-Labsch-2017}.  In particular, FUV photons originating from high-mass stars embedded within their natal molecular clouds with energies greater than 11.3\,eV are able to penetrate their surroundings by greater than parsec scales \citep{Howe93} to ionise atomic carbon in regions where hydrogen is mostly neutral \citep{SalgadoII}.  

\cite{Seon-2011} found a strong correlation of FUV radiation field with H$\alpha$ emission outside the bright H{\sc ii} regions due to a common radiative transfer mechanism.  Cooling of the gas in the PDR is dominated by C{\sc i} infrared line emission at 158\,$\mu$m, molecular rotational lines of CO (J=1$\rightarrow$0), and generates a mechanism to create carbon radio recombination lines (CRRLs) detectable in the radio part of the electromagnetic spectrum \citep{Sorochenko-2010,Andree-Labsch-2017}

The Orion molecular cloud complex is one of the closest regions of active high-mass star formation, making it ideal for the study of PDRs \citep{Gordon-RRL,ODell15}. The Orion molecular cloud was first modeled to have multiple layers of ionised gas by \cite{Ahmad-1976} and \cite{Boughton-1978} further elaborated on the ideas. \cite{Boughton-1978} concluded that within the H{\sc i} foreground region, carbon must be a significant source of the free electrons.  Both studies suggest that the low-frequency RRLs (n$<$190) arise in clouds associated with the H{\sc i} (21\,cm) component of the cold interstellar medium with temperatures $<$100\,K and gas densities $\approx$ 50\,cm$^{3}$.

Since then, a complex filamentary structure around Orion, on both large and small scales, has been widely observed and discussed \citep{Pon-2014,Pon-2016,Andree-Labsch-2017}.  Even early optical studies with high angular resolution \citep{Osterbrock-1959} suggest that the filamentary structure of the H{\sc ii} regions is on scales that would not be resolved with most telescopes and \cite{Gordon-RRL} suggest that many of the models have not incorporated these inhomogeneities, making interpretation of observational results difficult.  

Recent three-dimensional modeling, mixed with observational data by \cite{Andree-Labsch-2017}, of the PDRs around Orion show fractal structures that contain layers and explains why a large fraction of molecular material is located near the surface of the cloud. \cite{Pon-2014} identified linear H{\sc i} features extended radially away from the Orion Kleinmann$-$Low Nebula (Orion KL) with significant increases in regions where the H$\alpha$ decreases. \cite{Pon-2016} attributed this behavior to ionizing photons breaking out of the Orion-Eridanus superbubble surrounding the Orion region, extending about 180\,pc from our Sun.  

Observations of H$\alpha$ \citep{Finkbeiner-2003}, C{\sc i}, C{\sc ii}, CO (6$\rightarrow$ 5) \citep{Stutzki91, Howe93}, CO(1$\rightarrow$0) \citep{Dame-2001}, and CO(7$\rightarrow$ 6) \citep{Burgk89, Howe93} completed in and around Orion trace the complex regions of PDRs. \cite{Howe93} constrained the gas temperature of the Orion KL to greater than 40\,K and found the H$_{2}$ gas to have column densities exceeding 10$^{22}$\,cm$^{-2}$.

By using the Murchison Widefield Array (MWA) we are able to image and study a large fraction of the Orion molecular cloud complex, including both Orion A and Orion B, within a single field-of-view.  In this paper we present the lowest frequency detections (around 100\,MHz) of alpha and beta carbon radio recombination lines in the Orion region.  This work is done in conjunction with a molecular line survey published by Tremblay et al. (submitted), using a field-of-view of 400\,square\,degrees centred on the Orion KL Nebula.

\section{Observations \& Data Reduction}
Using the Murchison Widefield Array (MWA; \citealt{Tingay13}) at a centre frequency of 114.56\,MHz, we observed the Orion Nebula (RA(2000) = 05$^\text{h}$35$^\text{m}$, Dec.(2000) = $-$05$^\text{o}$27$^\text{'}$) for three hours on November 22, 2015 (see Table \ref{obs} for more details). The observations used snapshot imaging, which involve recording data in two minute intervals before adjusting the pointing. Pictor A (RA(2000) = 05$^\text{h}$19$^\text{m}$49.7$^\text{s}$, Dec.(2000) = $-$45$^\text{o}$46$^\text{'}$44$^\text{''}$, 452\,Jy at 160\,MHz) was observed for use as a primary calibrator at the start and end of the observation run. Each observation of Pictor A includes two minutes of data, and these data were used to calibrate tile bandpasses and phases. 

\begin{table}
\small
\caption{MWA Observing Parameters}

\label{obs}
\begin{tabular}{lc}
\hline
Parameter & Value\\
\hline
\hline
Central frequency& 114.56\,MHz\\
Total bandwidth & 30.72\,MHz\\
Number of imaged channels & 2400\\
Channel separation & 10 kHz (23\,km\,s$^{-1}$)\\
Synthesized beam FWHM & 3.2$^{\prime}$\\
Primary beam FWHM & 30\, degrees\\
Phase center of image (J2000) &  05h35m, --05d27m\\
Time on source & 3\,hours \\
						
\hline
\end{tabular}
\end{table}

The MWA has a contiguous bandwidth of 30.72\,MHz and using a two stage polyphase filterbank the data are channelised into 24$\times$1.28\,MHz `coarse' channels and each coarse channel is divided into 128$\times$10\,kHz `fine' spectral channels. This allows for simultaneous observations of multiple RRL transitions, including 34 $\alpha$--lines and 35 $\beta$--lines, as identified from the Splatalogue database \citep{Remijan-2007} . Even though the MWA offers 3072$\times$10\,kHz frequency spectral channels, only 2400 (100 fine channels of the 24 coarse channels) were imaged, in an effort to avoid imaging artifacts caused by the aliasing of the polyphase filter bank. Thus, only 78\,per\,cent of the bandpass was imaged, which could potentially detect 12 of the carbon $\alpha$--lines and 12 $\beta$-lines.

\subsection{Calibration \& Imaging}

The data calibration and imaging was completed as described in \cite{Tremblay17} for the molecular line survey of the Galactic Centre. At the observing frequencies used for these data, the MWA has a 900\,deg$^{2}$ full-width at half-maximum (FWHM) field-of-view (FOV) in the primary beam, but also has considerably large sidelobes. Bright sources in these sidelobes can corrupt observations by adding additional noise, and therefore need to be handled to correctly represent the desired radio data. To this end, we followed the approach used in the GaLactic and Extragalactic All-sky Murchison Widefield Array (GLEAM) survey \citep{GLEAM} to ``peel'' bright sources from the visibilities; in particular for these data, the Crab Nebula (RA(2000) = 05$^\text{h}$34$^\text{m}$34.94$^\text{s}$, Dec(2000) = 22$^\text{o}$00$^\text{'}$37.6$^\text{''}$; 1256\,Jy at 160\,MHz). 

Each two-minute observation experienced slightly different phase distortions due to the changing ionospheric conditions.  This manifested as small ($\approx$$20$\,arcsec) direction-dependent shift to the positions of sources in every observation.  A continuum image for each coarse channel of each observation is used to derive a single correction in R.A. and Dec. that is applied to each image cube.  After correction, the residual position offset is --1$\pm$17\,arcsec in R.A. and 5$\pm$17\,arcsec in Dec within the integrated image. To calculate the amount of image blurring of the point spread function, likely dominated by the ionosphere, the ratio of integrated flux density to peak flux density was computed and found to have an average value of 1.01$\pm$0.07, suggesting that we are robust against ionospheric activity for these observations. The average flux density error, when compared to the GLEAM catalog, was 0.1\%.  

The velocity resolution within the 99--129\,MHz frequency band ranged from 23 to 30\,km\,s$^{-1}$.  Doppler correction terms are currently not incorporated into the MWA imaging pipeline so the uncertainty of the velocity is 3.4\,km\,s$^{-1}$, significantly less than the channel width of the MWA.

\subsection{Survey Details}

A full description of the survey strategy and statistical analysis is presented in the Orion molecular line survey by Tremblay et al. (submitted).  In this paper, we provide a brief description.

The continuum-subtracted 10\,kHz spectral fine channels had a mean RMS of 0.4\,Jy\,beam$^{-1}$ at the phase centre.  Any channel having a RMS of $>$0.6\,Jy\,beam$^{-1}$ was flagged to remove channels effected by spurious RFI.  This reduced the search volume from the 2400 imaged channels to 1240 channels.  

The primary beam is the sensitivity pattern the telescope has on the sky.  Although 625\,square degrees is imaged (see Figure \ref{cont}) the search is limited to the most sensitive region of the primary beam using {\sc mimas} \citep{Hancock12} to reduce the search volume to $\approx$400\,square degrees (35\% reduction).

Each continuum-subtracted fine channel is independently searched using {\sc aegean} (\citealt{Hancock12,Hancock-2018}) to find pixels, in absorption and emission, with peak flux densities greater than 5\,$\sigma$ in comparison to an input RMS image.  The RMS image (as shown in in the upper right hand panel of Figure \ref{cont}) is the spectral RMS at each pixel position for each coarse channel (100 fine channels). The resultant catalog of potential sources was further reduced by removing signals that had a spectral RMS $>$~0.5\,Jy\,beam$^{-1}$ to reduce the chance of a false detection due to uncorrected continuum subtraction or image artifacts. 

The remaining sources in the catalog were further filtered.  Any source where the spectral RMS and image RMS were not within 20\% of each other, were removed from the catalog.  After this last stage of filtering on the catalog of sources from {\sc aegean} we found 8 positions on the sky, with 12 signals, that passed these criteria; of which 5 potential detections are presented within this work.

Using two-sided Gaussian statistics, we would expect no more than 78 signals with the 95000 independent synthesised beams and 1240 channels searched at a 5$\sigma$ threshold.  However, by filtering the data on quality to ensure the spectral RMS is $<$0.5\,Jy\,beam$^{-1}$ (85--90\% reduction in search volume) and to ensure that the spectral and image RMS were within 20\% of each other (74\% further reduction in search volume), the expected number of false positives due to thermal noise (in this reduced search volume) is less than one.  

We note here that {\sc aegean} fits Gaussians to the pixel data and applies a correction of the background to calculate the flux density for these potential detections.  The background is calculated as the 50th percentile of flux distribution in a zone 30 times the size of the synthesised beam.  However, to make the spectrum contained within this paper, the peak flux density in the spectrum is the flux density at the pixel position of potential detection as reported by {\sc aegean} but not the flux density reported by {\sc aegean}.  Therefore, it is possible that the flux densities shown in our spectra are underestimated, and consequently their significance is also underestimated. Our false positive rate may also be a slight overestimate as a result.

In the frequency range of 99-129\,MHz there are 273 known molecular and recombination line transitions within the 30.72\,MHz bandwidth.  Therefore, we would expect most of the 2400 imaged fine channels to be free of lines, as we observed.  The separation, in frequency, between the known atomic and molecular transitions is significant enough that incorrect identification of a detected signal is unlikely.  However, it may be possible that a recombination line overlaps with an unknown molecular transition and the identification we have made is incorrect. 

The chance that a noise signal being identified as an atomic or molecular line, that is within three fine spectral channels ($<$ 90\,km\,s$^{-1}$) of the rest frequency, is 0.4\%.  The chance of a significant noise peak being confused for a real detection, when multiple transitions are detected in a single location, is $<$0.1\%. 

\section{Results}
\subsection{Radio Continuum, H$\alpha$ and Dust Emission}
\begin{figure*}
		 
	 {%
	 \includegraphics[width=1\textwidth]{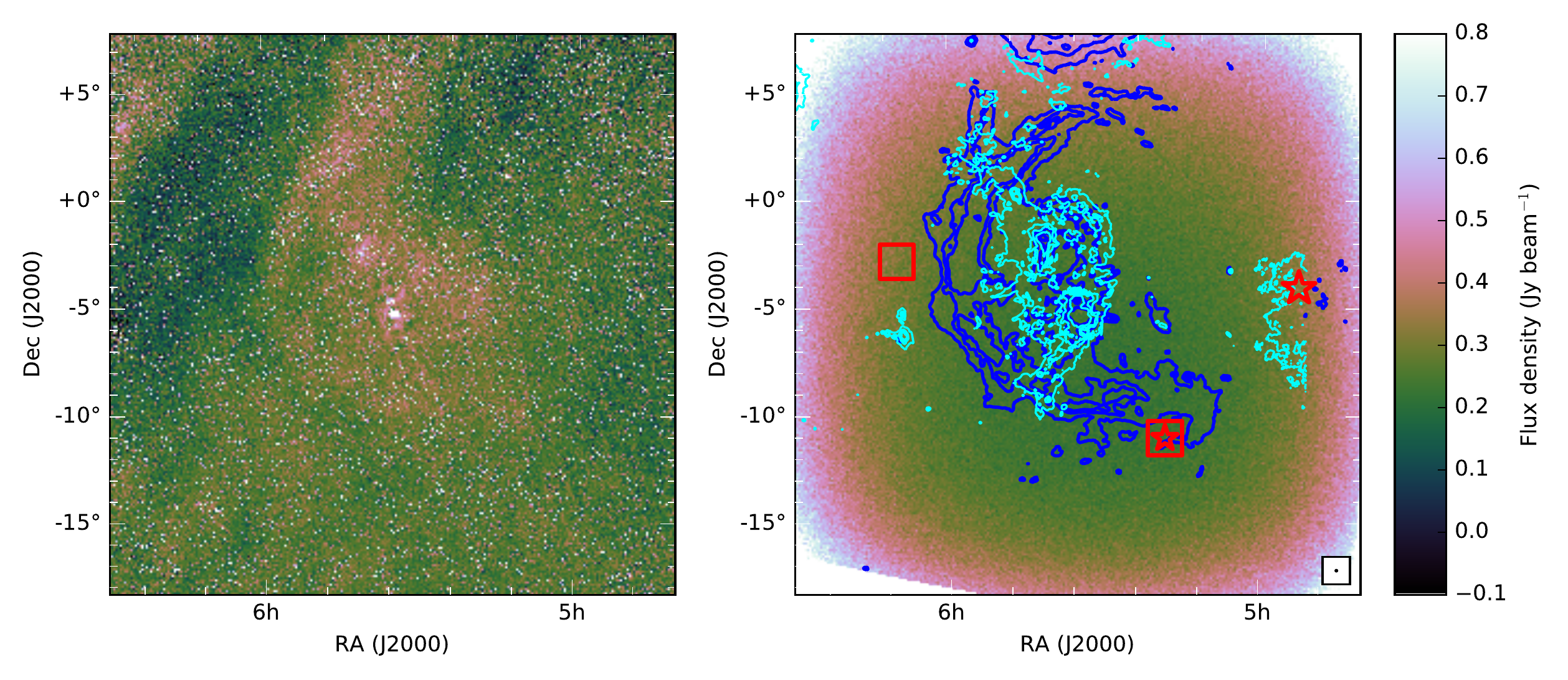}
	}
    
     {%
	 \includegraphics[width=0.99\textwidth]{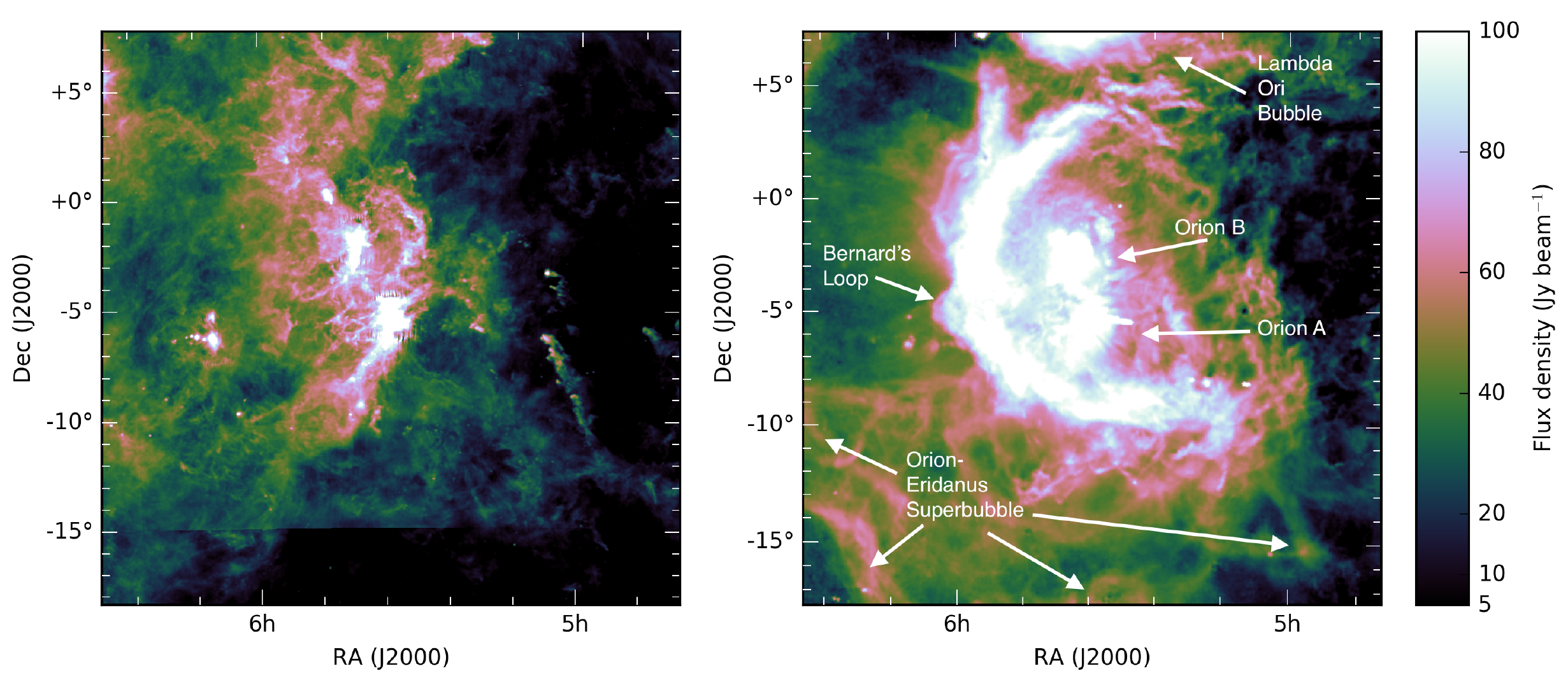}
	}
   	\caption{The upper left image is a continuum image of the Orion survey region across the 30.72\,MHz of bandwidth with a central frequency of 114.6\,MHz showing the full region blindly searched for molecular signatures and recombination lines.  The upper right hand image is an RMS map showing the sensitivity of the MWA in a typical coarse channel data cube.  The C393$\alpha$ and C496$\beta$ transitions are represented by squares and stars respectively.  The contours in cyan trace the image in the bottom left, which shows dust emission at 100\,$\mu$m surveyed by \cite{Schlegel}. The blue contours and the image in the bottom right are the optical H${\alpha}$ emission from the Southern H-Alpha Sky Survey Atlas (SHASSA) \citep{Finkbeiner-2003}.  We note that there is an image artifact in the 100\,$\mu$m survey data on the bottom, created by combining the data from different fields.}
\label{cont}
\end{figure*}

A continuum image from these observations centered on the Orion Nebula at 114.56\,MHz is shown in the top left image in Figure \ref{cont}.  The angular resolution of the 114\,MHz image obtained from the MWA is 73$^{\prime\prime}$ x 67$^{\prime\prime}$ and the total flux density in the Orion Nebula is 45$\pm$4\,Jy. 

The top right hand image of Figure \ref{cont} shows the H$\alpha$ emission at 653\,nm \citep{Finkbeiner-2003} in blue contours overlaid on a RMS map for the coarse channel data cube at 103\,MHz, which is representative of the typical RMS map.  The RMS map shows the sensitivity pattern on the sky across the 625\,square degrees FOV imaged in this survey.  

The H$\alpha$ emission, shown in the bottom right hand side of Figure \ref{cont} was observed by the Southern H-Alpha Sky Survey Atlas (SHASSA) survey and traces the Bernard's loop, the Lambda Ori bubble (supernova remnant), and the Orion-Eridanus filaments. Bernard's loop is a photoionised region that is particularly interesting as theoretical models and observations only agree if the region is enhanced with heavy metals \citep{odell_ferland_porter_hoof_2011}. It is suggested that the crescent shape is caused by an old supernova remnant that, as the bubble expanded, has swept up dust and gas as it moved through the high density and pressure gradients associated with the Galactic gas layer \citep{Wilson-2005,Pon-2016}.  

Within the continuum image of our survey, on the top left, the faint synchrotron radiation is correlated with the H$\alpha$ emission surrounding the nebula.  The cyan contours on the RMS map and the image in the bottom left, trace the dust emission at 100\,$\mu$m \citep{Schlegel} and the tentative detections are mostly contained in the region of the cloud complex where the dust and H$\alpha$ emission is decreased and where the H{\sc i} density is expected to increase \citep{Pon-2014}.  

\subsection{Carbon Recombination Lines}
Carbon recombination lines C393$\alpha$ and C496$\beta$ are tentatively detected within the Orion molecular cloud complex, as shown in the spectra in Figure \ref{C393}. These detected transitions appear to be co-located with H$\alpha$ filamentary structure and in regions where the 100\,$\mu$m dust density is decreased.   

The derived parameters for each source are found in Table \ref{alpha} and \ref{beta}.  The Rayleigh-Jeans approximation is used to determine the brightness temperature (T$_{\mathrm{B}}$):

\begin{equation}
T_{\mathrm{B}}=\frac{\lambda^{2}}{2k\Omega}\mathrm{S},
\end{equation}

where $\lambda$ is the wavelength of the recombination line, $k$ is the Boltzmann constant, $\Omega$ is the beam solid angle, and S is the flux density in Jy\,beam$^{-1}$.  Since the signal is not resolved spectrally, the integrated intensity is the brightness temperature integrated over the velocity resolution of the single channel. 

\begin{equation}
\int\mathrm{Intensity} = S \times \mathrm{v}
\end{equation}

where v is the velocity resolution of the channel of the detected recombination line. The calculated brightness temperature of $\approx$4000\,K for the sources is not unexpected, as low-frequency transitions have a stimulated emission component by the domination of collisional excitation \citep{SalgadoII}. The temperature of the background continuum (T$_{c}$) is largely uniform over the FOV and is calculated as 1100\,K scaled from the Haslam model \citep{Remazeilles} at 408\,MHz using a spectral index of $-$2.6.

\begin{figure*}
	\centering    	 
	 {%
	 \includegraphics[width=0.8\textwidth]{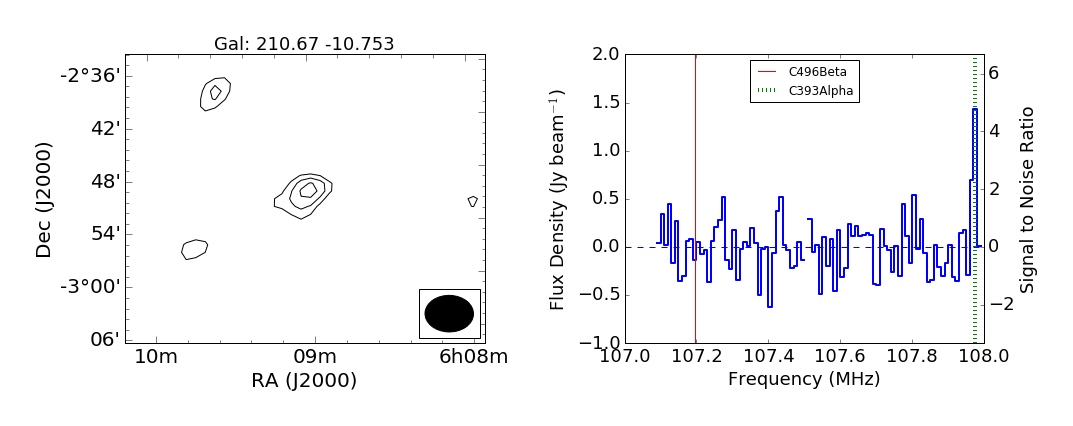}
	}
    \centering    	 
	 {%
	 \includegraphics[width=0.8\textwidth]{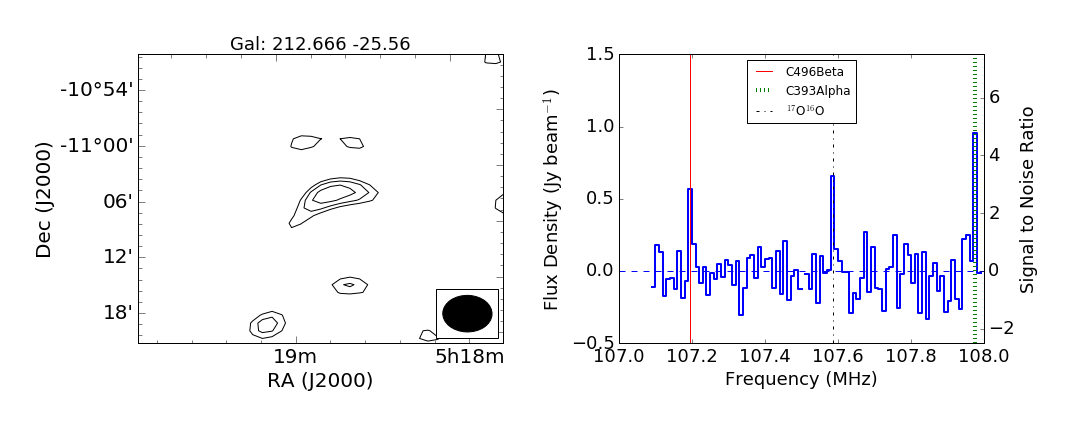}
	}
    	\centering    	 
	 {%
	 \includegraphics[width=0.8\textwidth]{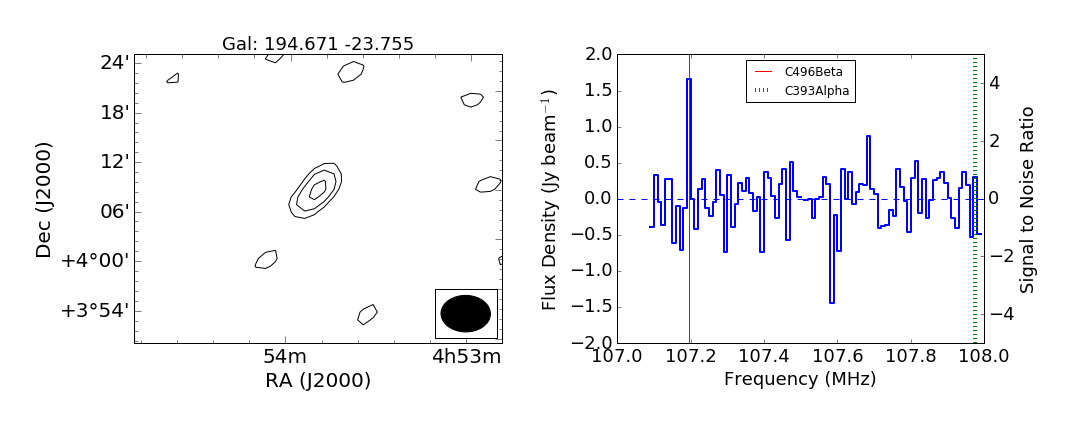}
	}
	\caption{A tentative detection of the C393$\alpha$ (107.98\,MHz) and C496$\beta$ (107.19\,MHz) at the Galactic coordinates listed on the top of each plot.  The contours plot on the left represent the 3, 4, 5, and 6\,$\sigma$ levels of the associated detection. The spectrum on the right is representative of the spectrum of the coarse channel band in which the tentative detection was made with the flagged channels blanked. The spectra at G212.67~$-$25.56 also shows a possible detection of C496$\beta$ at 107.18\,MHz and molecular oxygen ($^{17}$O$^{18}$O) at 107.60\,MHz.}
\label{C393}
\end{figure*}

\begin{table}
\small
\caption{Information about the tentative detections of C393$\alpha$.  The peak-pixel positions of each tentative detection is listed in Galactic coordinates (source), as well as Right Ascension (R.A.) and Declination (Dec.).  The flux density, velocity, integrated intensity and brightness temperature are provided for each location the transition was detection.}

\label{alpha}
\resizebox{\columnwidth}{!}{%
\begin{tabular}{llcc}
\hline
	&	&C393$\alpha$		&	\\
\hline
\hline
Source	&&	G210.67 $-$10.75	&	G212.67 $-$25.56\\							
R.A.	&	\small(J2000)		&6:09:02&	5:18:48	\\
Dec.	&	\small(J2000)	&		$-$2:53:56&	$-$11:07:41	\\
Flux &&&\\
Density & \small(Jy\,beam$^{-1}$) & 1.43 & 0.97\\
V$_{\mathrm{lsr}}$	&	\small(km\,s$^{-1}$)	&0$\pm$24&	0$\pm$24	\\
$\int$Intensity	&	\small(K km\,s$^{-1}$)	&9.9$\times$10$^{4}$&	9.5$\times$10$^{4}$	\\
T$_{\mathrm{B}}$	&	\small(K)		&4136	&4107		\\						
\hline
\end{tabular}%
}
\end{table}

\begin{table}
\small
\caption{Information about the tentative detections of C496$\beta$.  The peak-pixel positions of each tentative detection is listed in Galactic coordinates (source), as well as Right Ascension (R.A.) and Declination (Dec.).  The velocity, integrated intensity and brightness temperature are provided for each location the transition was detected.}

\label{beta}
\resizebox{\columnwidth}{!}{%
\begin{tabular}{llll}
\hline
&	&	C496$\beta$			&	\\
\hline
\hline
Source	&		&	G212.67 $-$25.56	&	G194.67 $-$23.76	\\				
R.A.	&	(J2000)	&5:18:48&		4:53:56\\
Dec.	&	(J2000)	&	$-$11:07:41&		+04:06:13.42\\
Flux &&&\\
Density & (Jy\,beam$^{-1}$) & 0.97 & 1.72\\
V$_{\mathrm{lsr}}$	&	(km\,s$^{-1}$)&	24$\pm$24  &24$\pm$24\\
$\int$Intensity	&	(K km\,s$^{-1}$)	&		6.7$\times$10$^{4}$&	12$\times$10$^{4}$\\
T$_{\mathrm{B}}$	&	(K)		&2805	&4974	\\						
\hline
\end{tabular}%
}
\end{table}

The full spectrum from the position of each of the potential detections is shown in the Appendix.  This shows that no other transitions of carbon alpha or beta lines were detected.  However, at the position of G212.67 $-$25.56 the data cube contains a signal around 5\,$\sigma$ associated with the rest frequency of C393$\alpha$ and also contains two emission peaks around 3\,$\sigma$ that are associated with the rest frequency of the C496$\beta$ line at 107.18\,MHz and the rest frequency of molecular oxygen ($^{17}$O$^{18}$O) at 107.60\,MHz.  The  upper energy level divided by the Boltzmann constant for the transition of molecular oxygen at 107.60\,MHz is 39.81\,K and represents the N= 4--4, J=3--3, F=$\frac{7}{2}$--$\frac{5}{2}$ transition.  The same transition of molecular oxygen is possibly detected at the position of G194.471 $-$23.755 in absorption (Figure \ref{C393}) and other locations reported in the molecular survey paper (Tremblay et al., submitted).

\section{Discussion}
\subsection{Turn-over Frequency of CRRL Sources}

Observations of CRRLs have spanned frequencies of 10 to over 5000\,MHz \citep{Payne94,Roshi,Salas17}. \cite{Peters2010} summarised the information about known Galactic low-frequency CRRLs and, using a modified version of the radiometer equation, calculated the approximate amount of observing time required in order for the new generation of low-frequency telescopes to detect them.  The MWA was not on the list, but using the same equation we could expect to detect them in the averaged spectra in $\approx$ 2\,hours if all the observed lines are in either absorption or emission.  However, due to the population effects, the exchange from absorption to emission will likely occur between 100 and 150\,MHz, making this a difficult experiment.

At large n--bound states, observed at frequencies less than 100\,MHz, the relative populations of atomic levels are controlled primarily by collisional processes.  This makes the population levels close to the kinetic temperatures of less than 100K and so the lines are detected in absorption along the line of sight toward strong continuum sources.  At lower n--bound states, observed at frequencies greater than 200\,MHz, the relative populations are dominated by radiative processes; moving the excitation temperature away from the kinetic temperature towards negative values.  The level populations then become inverted and the recombination lines are detection in emission.  This happens even if the brightness temperature is higher than the kinetic temperature \citep{Payne89,SalgadoI}.

The transition between emission and absorption, often defined as the position where the optical depth is between 1 and 1.5 \citep{Gordon-RRL}, was first modeled by \cite{Shaver75} and later observed by \cite{Payne89} in Cassiopeia A. The transition between absorption and emission must occur regardless of the details of the population mechanism but the actual turnover frequency and the strength of the emission lines will depend on the density and temperature of the clouds being observed. 

\cite{Gordon-RRL} explain this in terms of the Bohr atom.  At low radio frequencies, the size of the atoms is large, making them more likely to interact with the charged particles in the diffuse H{\sc ii} gas.  The wavelength of these turnover frequencies is directly related to the probabilities of these collisions happening, such that, as the gas densities increase the rate of the collision decreases. So, in dense gas the turn-over frequency is likely to be at higher observational frequencies than in diffuse gas.  

Therefore, the observation of CRRLs in the frequency range of 100$-$200\,MHz is an important probe toward understanding the cold neutral medium around PDRs and H{\sc ii} regions.  The observed emission of the alpha and beta lines reported here at 107--108\,MHz, suggest the turnover is at frequencies less than 107\,MHz for the ionisation layer  and, as expected, the transitions are observed in emission within the regions of diffuse gas.

\cite{Payne89} observations of the turn-over between emission and absorption within Cassiopeia A demonstrated that in between these two states, the spectra are flat and recombination lines are undetectable.  That may explain why we are not detecting signals associated with CRRLs below 107\,MHz but above 108\,MHz we would expect to detect the other known transitions in emission. Historically, due to various instrumental and population effects, not all expected transitions are observed (e.g. \citealt{Bell97}) or the observations were focused on a single transition (e.g. \citealt{Wyrowski97}), both of which is not unexpected within the spectral line community \citep{Herbst09}. However, we will explain possible reasons for the discrepancy within our data. 

\subsection{Spectral Quality}

Each fine (10\,kHz) spectral channel is individually imaged for each of the two-minute snap-shot observations and then built into an image cube for a single coarse channel. For a single coarse channel, all the observations are integrated together using inverse variance weighting.  Therefore each integrated image cube has unique sensitivities in comparison to any other coarse channel cubes within the observation set.  Also, due to the polyphase filterbank that channelizes the MWA data, each fine spectral channel has a steep bandpass shape, such that any detected signal that may not be centered within the 30\,km\,s$^{-1}$ resolution, may be lost.

These tentative detections are of low significance ($\approx$5$\sigma$).  The low significance, combined with the knowledge of the data processing and characteristics of the MWA described above, each spectral cube for each coarse channel has different noise properties.  This suggests that it is not improbable that all the transitions for the same atom are not detected.  Therefore, we call these detections tentative as they are unresolved and the subsequent, unflagged, transitions within our band were not detected.   

\subsection{Origin of CRRLs around Orion}

Modeling and observations from \cite{Ahmad-1976} and \cite{Boughton-1978} introduced the idea that the Orion complex contains multiple layers of ionised gas. \cite{Boughton-1978}, and later confirmed by  \cite{Wyrowski97}, determined that the carbon around Orion is ionised in a layer that is between the observer and the nebula and not within the nebula itself. This is consistent with other observations within the region where most emission is identified from a region of ionised gas on the observer's side of the Orion Molecular cloud and in a PDR beyond the primary ionisation front \citep{ODell15}.  

Theoretical modeling by \cite{Pon-2016} suggests that the Orion-Eridanus super-bubble is a layer of expanding ionised gas with the outer regions about 180\,pc from the Sun.  They also determined that the total ionizing luminosity of the Orion star-forming region is sufficient to produce the H$\alpha$ emission observed within this super-bubble.  

Reported here are the lowest frequency spectral line observations of the Orion region published to date, as far as we are aware. With the resolution of the MWA it is likely these recombination line detections are within the diffuse filamentary structure of the near-field ionisation layer.  However, the frequency resolution of the MWA does not allow for analysis of the collisional and radiation broadening effects on the spectral line.  Therefore, no attempt to place upper limits on the electron temperature are made here.

\cite{Ahmad-1976} discussed the discrepancy of velocities at frequencies greater than 5\,GHz versus observations of CRRLs at frequencies less than 5\,GHz.  He suggested that the low-frequency observations are in cold gas in front of the H{\sc ii} region while carbon recombination lines at higher frequencies are detecting warm regions behind the H{\sc ii} region. 

Typical velocities for carbon recombination lines detected in the Orion KL, Orion A, and Orion B portions of the giant molecular cloud complex range from $\approx$--9\,km\,s$^{-1}$ \citep{Ahmed74, Wyrowski97,Roshi} at frequencies greater than 5\,GHz and $\approx$--6\,km\,s$^{-1}$ at frequencies down to 600\,MHz \citep{Chaisson-1974}.  However, \cite{Anantharamaiah-1990} detected carbon and hydrogen recombinations lines in the southern edge of Orion B with velocities of $\approx$ 48\,km\,s$^{-1}$.

Observations of the carbon monoxide J=1$\rightarrow$0 transition around the Orion molecular cloud trace the complex structure of carbon within the ionised frontal layer.  \cite{Wilson-2005} observed velocities ranging from $-$1 to 18.5\,km\,s$^{-1}$ with knots of compressed gas in regions of active star-formation, suggesting that there is a large gradient of possible velocities for carbon atoms within the Orion ionisation front.

The tentative detections of C393$\alpha$ and C496$\beta$ lines have velocities of 0$\pm$24\,km\,s$^{-1}$ and 24$\pm$ 24\,km\,s$^{-1}$ respectively which are consistent with previous CRRL observations within the region. However, due to the velocity resolution of the MWA we can not discriminate between the previously observed velocities within the different ionisation layers.    

\section{Conclusions}
We have tentatively detected one alpha transition and one beta transition across three positions within the Orion molecular cloud complex ionisation layer in front of the H{\sc ii} region, at a level greater than 5\,$\sigma$.  The transition between emission and absorption, at frequencies between 100--200\,MHz are important to identify and study the cloud dynamics.  However, due to the spectral resolution of the MWA we can not provide analysis of the collisional and radiative broadening effects. We can conclude that the transition within these regions is less than 107\,MHz as the signals detected in emission are at frequencies greater than this and the detected regions correspond to hot-spots of the diffuse hydrogen alpha emission observed with optical telescopes.

\begin{acknowledgements}
The authors would like to thank the referee for their useful comments and feedback. CDT would like to thank Paul Hancock with help with the coding used in data reduction. The authors would like to acknowledge the contribution of an Australian Government Research Training Program Scholarship in supporting this research. This work was supported by resources provided by the Pawsey Supercomputing Centre with funding from the Australian Government and the Government of Western Australia.  This scientific work makes use of the Murchison Radio-astronomy Observatory, operated by CSIRO. We acknowledge the Wajarri Yamatji people as the traditional owners of the Observatory site. Support for the operation of the MWA is provided by the Australian Government (NCRIS), under a contract to Curtin University administered by Astronomy Australia Limited.  We gratefully acknowledge the support of NASA and contributors of SkyView surveys.
\end{acknowledgements}

\bibliographystyle{pasa-mnras}
\bibliography{research3}

\appendix
\section{\\Full Spectra}
We present the full spectra from the MWA in the band of 99--122\,MHz to show the quality of the data.  The spectra represents the data in the positions of all three tentative detections.

\renewcommand\theContinuedFloat{\alph{ContinuedFloat}}

\begin{figure*}
  \ContinuedFloat*
  \includegraphics[width=\textwidth]{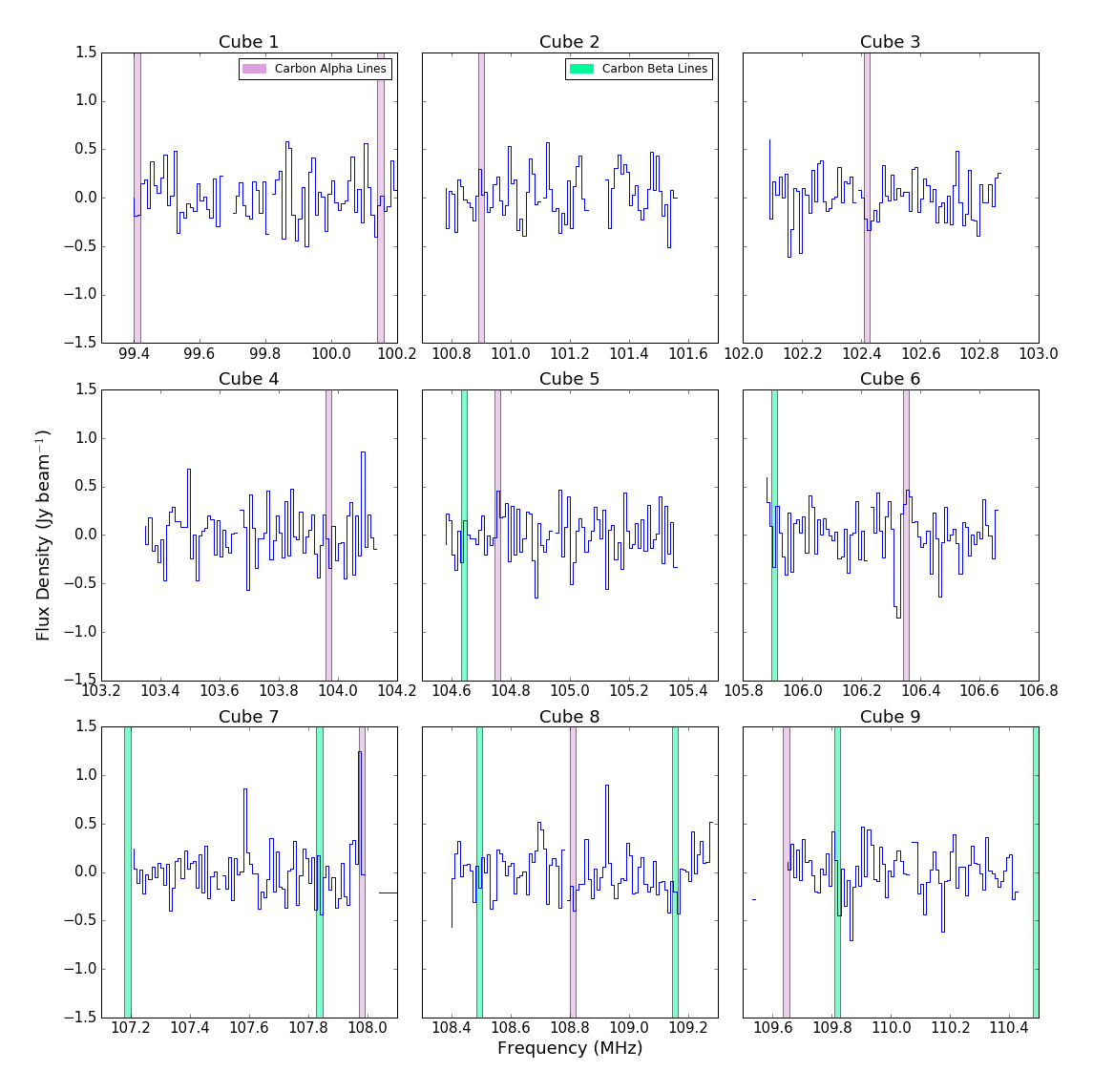}
  \caption{\label{first}MWA spectra for a data cube used within this survey at the position of G212.67 $-$25.56.  The alpha recombination rest frequency are marked in plum and the beta rest frequency positions are marked in green.  Flagged channels are blanked out in the spectra.} 
\end{figure*}
\begin{figure*}
  \ContinuedFloat
  \includegraphics[width=\textwidth]{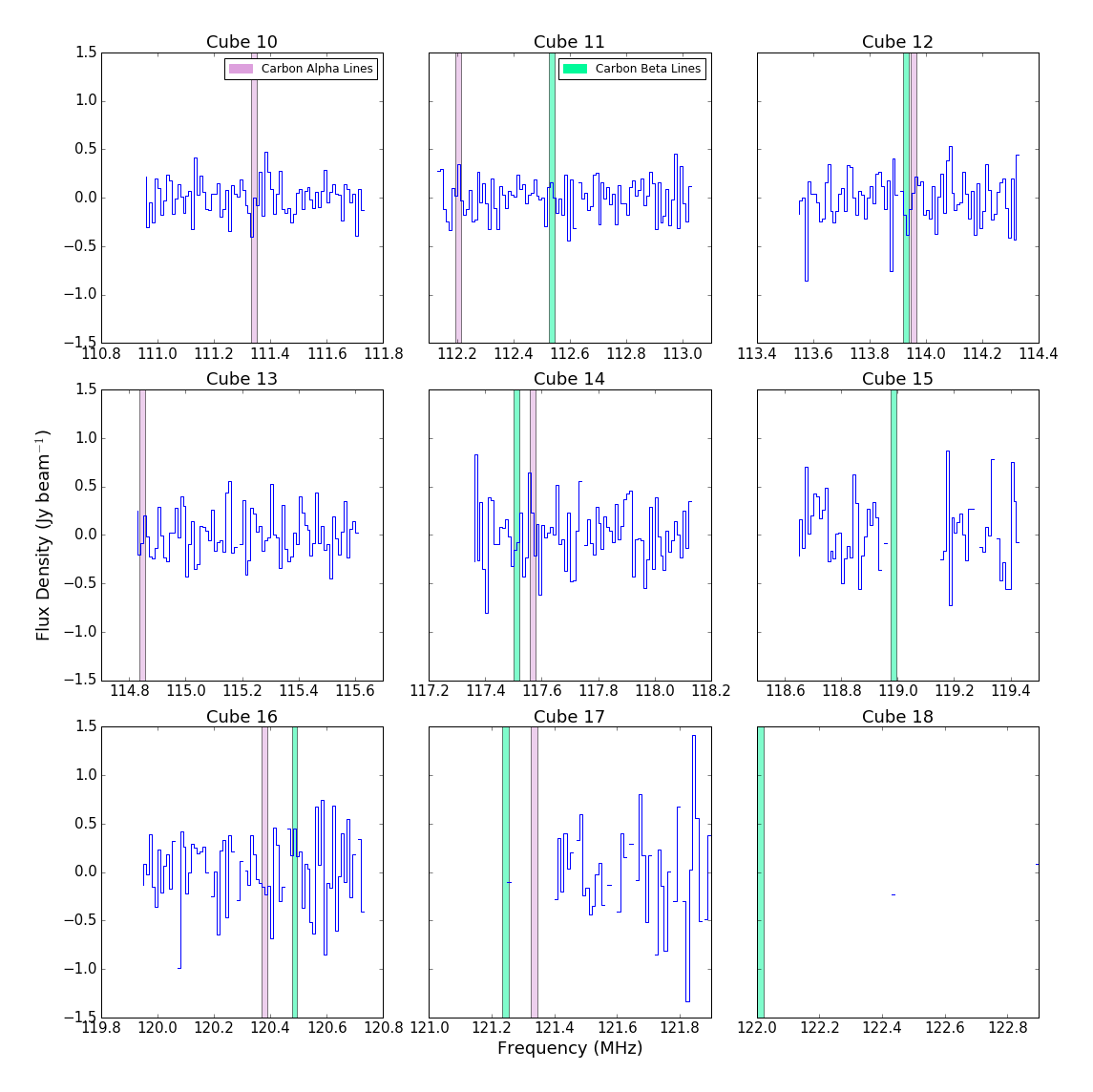}
  \caption{\label{second}MWA spectra for a data cube used within this survey at the position of G212.67 $-$25.56.  The alpha recombination rest frequency positions are marked in plum and the beta rest frequency positions are marked in green. Flagged channels are blanked out in the spectra.} 
\end{figure*}

\begin{figure*}
  \ContinuedFloat*
  \includegraphics[width=\textwidth]{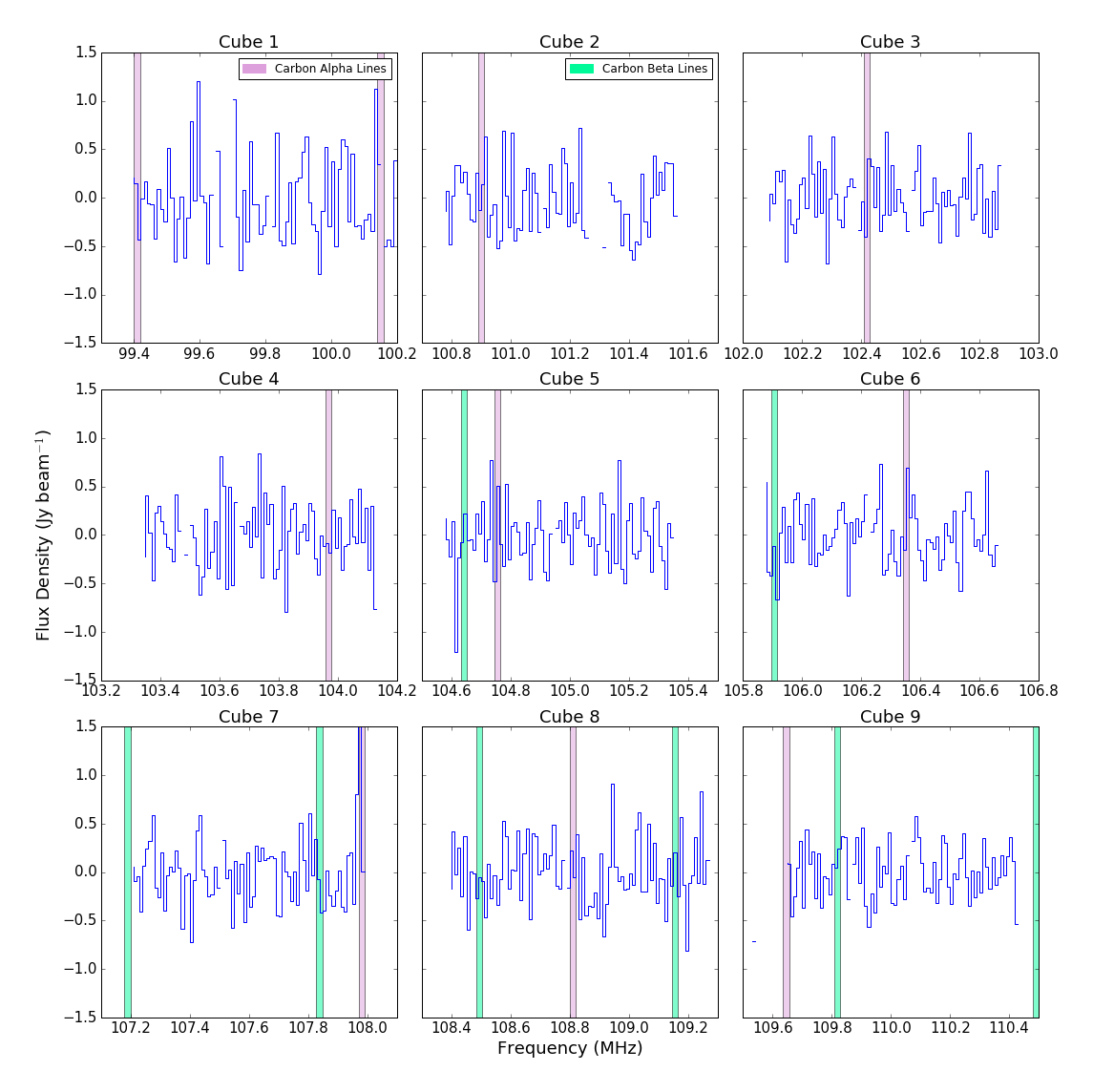}
  \caption{\label{first}MWA spectra for a data cube used within this survey at the position of G194.67 $-$23.76.  The alpha recombination rest frequency are marked in plum and the beta rest frequency positions are marked in green.  Flagged channels are blanked out in the spectra.} 
\end{figure*}
\begin{figure*}
  \ContinuedFloat
  \includegraphics[width=\textwidth]{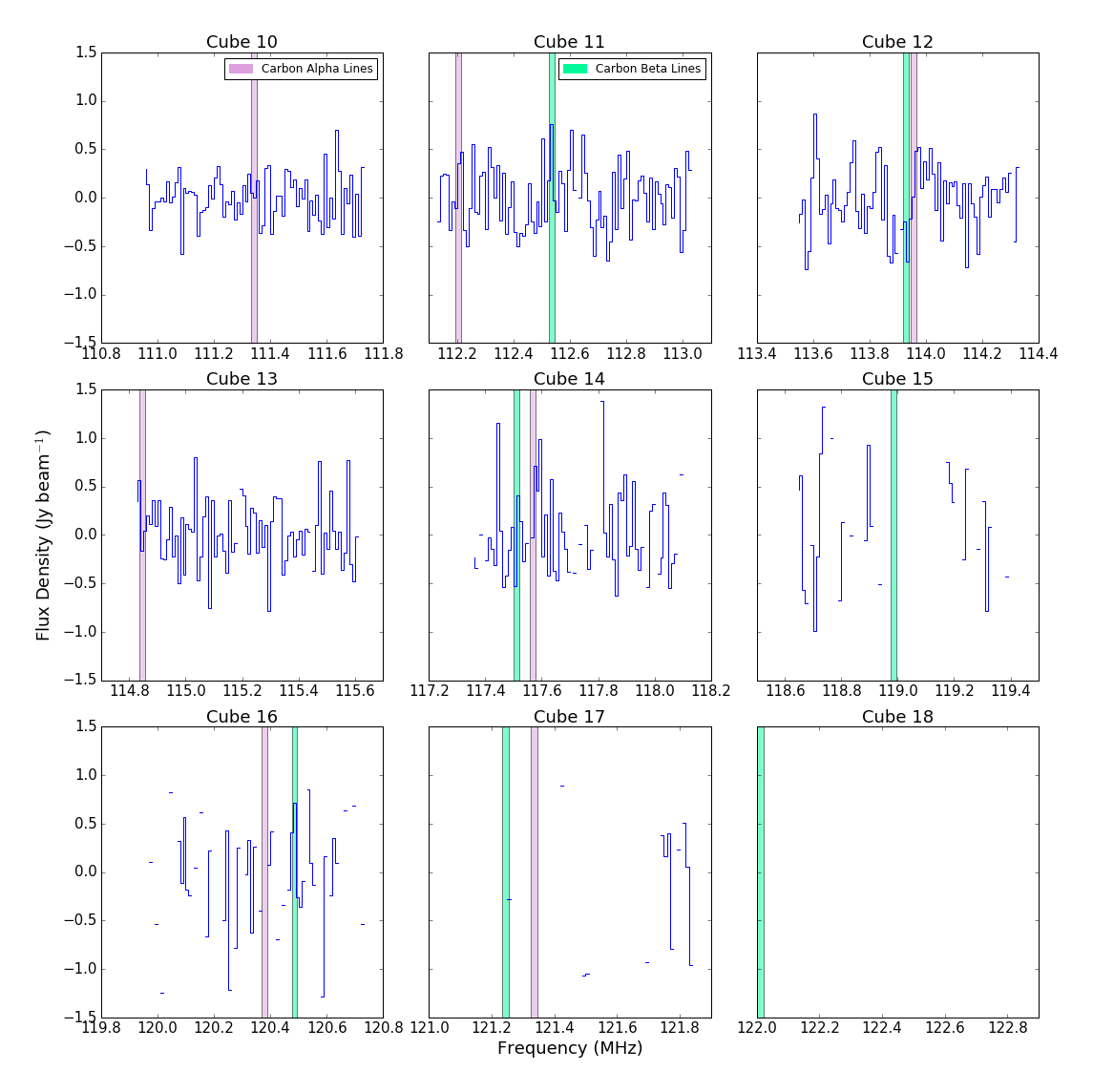}
  \caption{\label{second}MWA spectra for a data cube used within this survey at the position of G210.67 $-$10.75.  The alpha recombination rest frequency positions are marked in plum and the beta rest frequency positions are marked in green. Flagged channels are blanked out in the spectra.} 
\end{figure*}

\begin{figure*}
  \ContinuedFloat*
  \includegraphics[width=\textwidth]{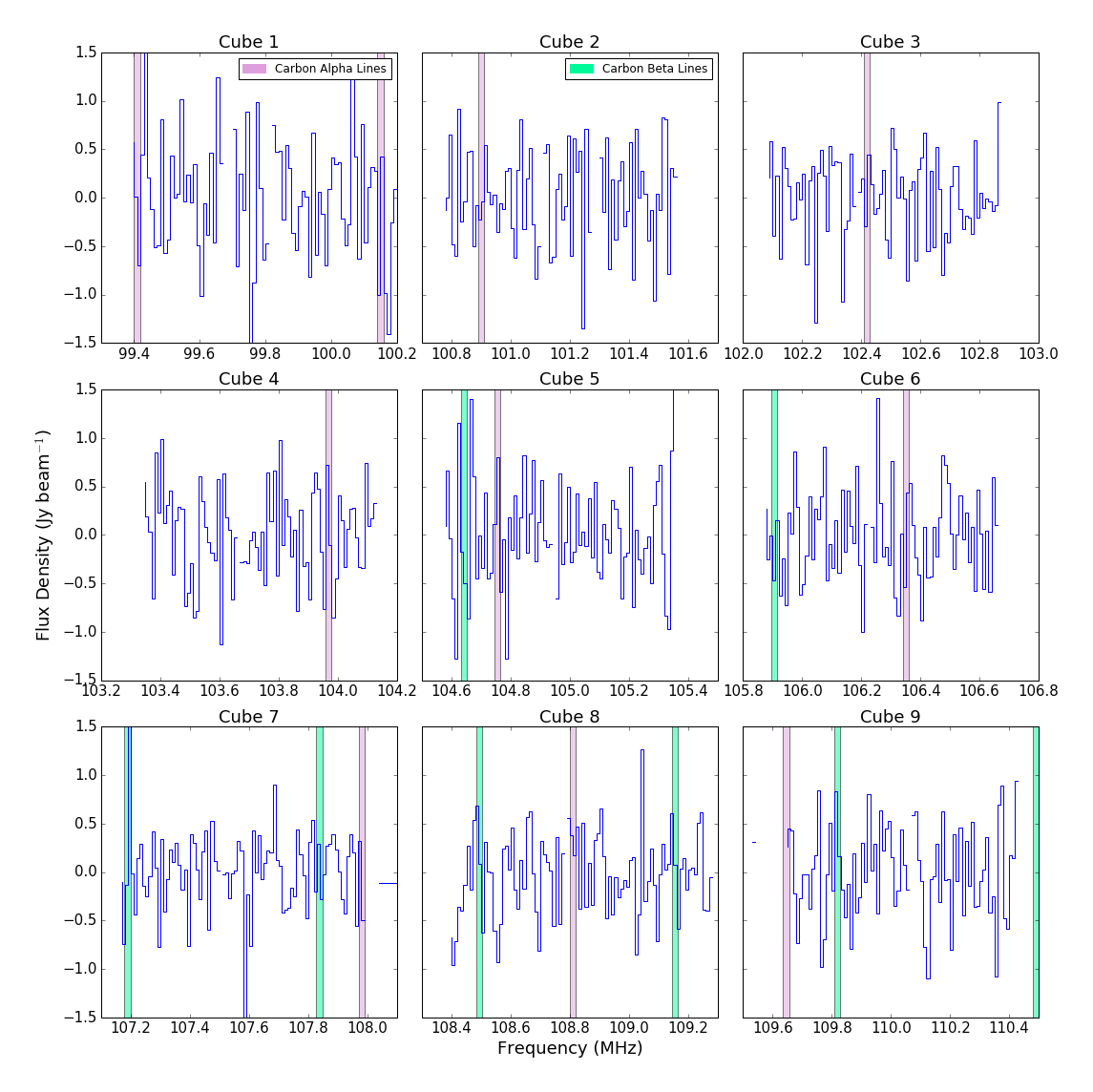}
  \caption{\label{first}MWA spectra for a data cube used within this survey at the position of G194.67 $-$23.76.  The alpha recombination rest frequency are marked in plum and the beta rest frequency positions are marked in green.  Flagged channels are blanked out in the spectra.} 
  
\end{figure*}

\begin{figure*}
  \ContinuedFloat
  \includegraphics[width=\textwidth]{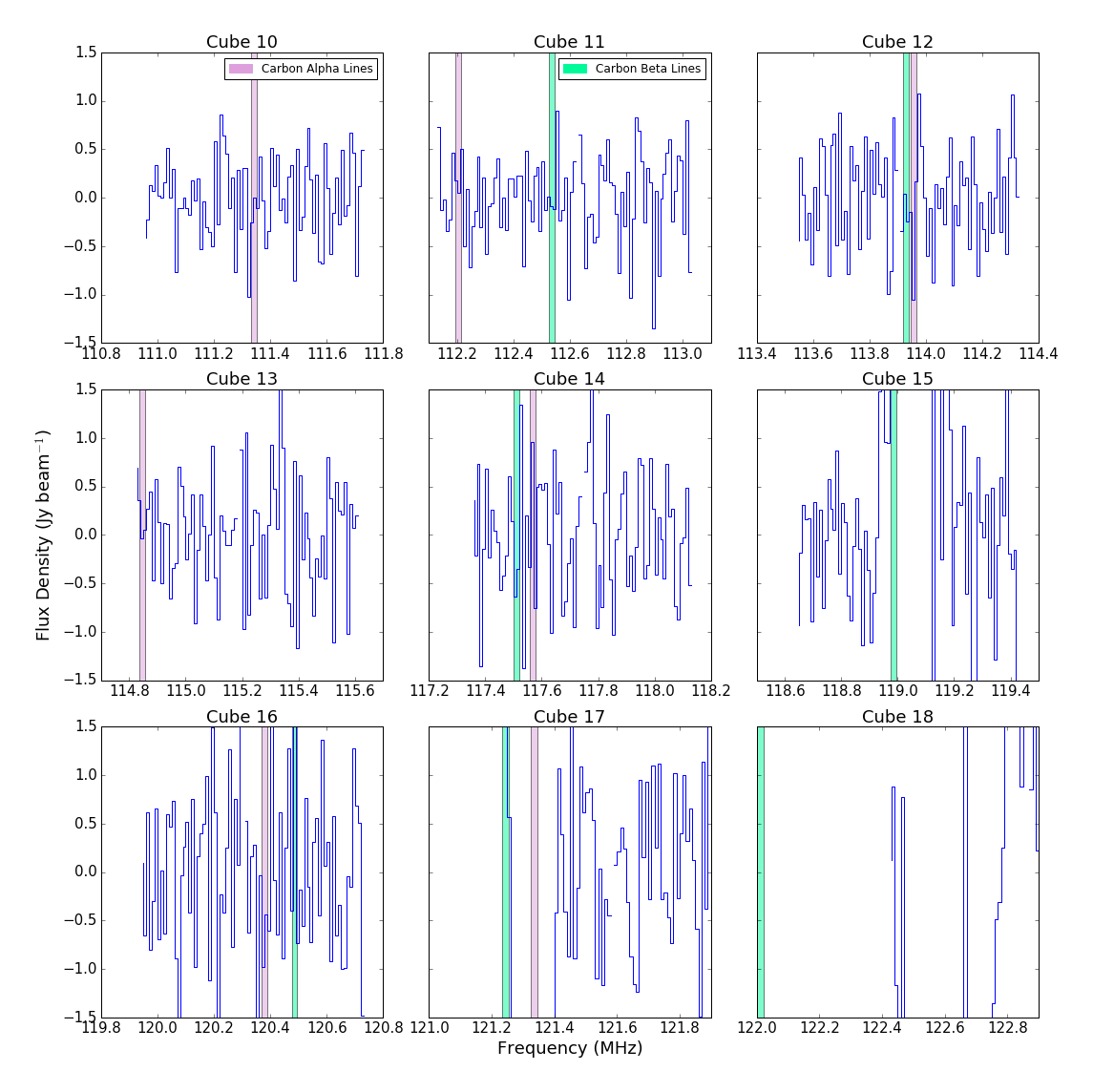}
  \caption{\label{second}MWA spectra for a data cube used within this survey at the position of  G194.67 $-$23.76.  The alpha recombination rest frequency positions are marked in plum and the beta rest frequency positions are marked in green. Flagged channels are blanked out in the spectra.} 
\end{figure*}

\end{document}